# An embryo of protocells: The capsule of graphene with selective ion channels

Zhan Li [1,2,†,*], Chunmei Wang[3,†], Longlong Tian[4,†], Jing Bai[2], Yang Zhao[5], Xin Zhang[2], Shiwei Cao[2], Wei Qi[4], Hongdeng Qiu[1], Suomin Wang[7], Keliang Shi[4], Youwen Xu[8], Zhang Mingliang[1],Bo Liu[4], Huijun Yao[2] , Jie Liu[4], Wangsuo Wu[4,*], Xiaoli Wang[3]

1, Key Laboratory of Chemistry of Northwestern Plant Resources, Key Laboratory for Natural Medicine of Gansu Province, Lanzhou Institute of Chemical Physics, Chinese Academy of Sciences, Lanzhou, 730000, P.R. China

2, Institute of Modern Physics, Chinese Academy of Sciences, Lanzhou, 730000, P.R. China

3, Lanzhou Institute of Husbandry and Pharmaceutical Sciences of CAAS, Lanzhou, 730000, P.R. China

4, Radiochemistry Laboratory, Lanzhou University, Lanzhou, 730000, P.R. China

5, Department of Chemistry, State Key Lab of Molecular Engineering of Polymers, Shanghai Key Lab of Molecular Catalysis and Innovative Materials, and Collaborative Innovation Center of Chemistry for Energy Materials, Fudan University, Shanghai 200433, P. R. China.

6, School of Nuclear Science and Technology, Lanzhou University, Lanzhou, 730000, P.R. China

7, Key Laboratory of Grassland Agro-ecosystem, Ministry of Agriculture, School of Pastoral Agriculture Science and Technology, Lanzhou University, Lanzhou 730000, P.R. China

8, Brookhaven National Laboratory Medical Department, Building 901, Room 106, Upton, NY 11973

[†]These authors contributed equally to this work.

[*]Correspondence should be addressed to: lizhancg@gmail.com; wuws@lzu.edu.cn.

**Abstract**

Many signs indicate that the graphene could widely occur on the early Earth. Here, we report that coating graphene oxides with phospholipid would curl into capsules in saturated solution of heavy metal salt of pH 0.15 at 4 $^{o}$C similar to the primitive ocean conditions. L-animi acids exhibit higher reactivity than D-animi acids for graphene oxides. Moreover, monolayer graphene with nanopores prepared by unfocused $^{84}Kr^{25+}$ has high selectivity for permeation of the monovalent metal ions ($Rb^+$ > $K^+$ > $Cs^+$ > $Na^+$ > $Li^+$), but does not allow $Cl^-$ through. It is similar to $K^+$ channels, which would cause efflux of some ions from capsule of graphene oxides with the decrease of pH in the primitive ocean, creating a suitable inner condition for origin of life. Interestingly, the specific structure of graphene tubes might be a synthetic template of nucleotides. We conferred that graphene might bred primitive life, and then evolved gradually.

## Introduction

The origin of life has always been an interesting and important topic. Its resolution could not only satisfy curiosity of people about this central existential issue, but could also reveal a directly related topic - the nature of the physic - chemical relationship linking living and nonliving matter[1]. However, the precise details of the origins for life remain elusive. A theory is generally accepted that current life on Earth descends from an RNA world[2,3]. However, it is an issue how the first RNA could transform its own catalyst (a ribozyme) and where did it happened? In 2004, Simon Nicholas Platts[4] proposed a new polycyclic aromatic hydrocarbons (PAH) world hypothesis to try to answer the first question. The world hypothesis of PAH proposes that PAH, known to be abundant in the universe[5,6], and assumed to be abundant in the primordial soup of the early Earth, played an important role in the origin of life by mediating the synthesis of RNA molecules, leading into the RNA world[4], but untested not yet. Moreover, another important issue is where did it happen, whether answers could also be found from the world hypothesis of PAH?

As we all know, the environment of primitive oceans was very harsh with a strongly acidic solution and containing a high concentration of heavy metal salt[7]. For this reason, identifying a suitable space would be indispensable for the protection of biomolecules, leading to efficient biochemical reactions[8]. Many scientists thought that protocell membranes play a protective role in the formation of RNA, and studied the origin of protocell membranes in detail. They proposed that fatty acids, with their constituent alcohols and glycerol monoesters, are attractive candidates as components of these cells[9]. However, Mulkidjanian et al.[10] discovered that $Na^+$ is not conducive to the synthesis of protocell membrane consisting of fatty acids and proteins. So, they thought that unless the first cell has ion-selective channels, the oceans could not contain the necessary balance of ingredients to foster life. This assertion was an enormous challenge to the classic theory that contended that the origin of life began in the oceans. However, we found that the introduction of graphene (belongs to PAH) would reconcile theories of Mulkidjanian not in favor of the classic one.

Researches indicate that graphene could be synthesized easily from methane or carbon dioxide, the main composition of the primitive reducing atmosphere[11], through lightning in the absence of oxygen[12,13], meaning that graphene is extremely common and abundant in in the primitive ocean in the early Earth. Furthermore, Sint et al.[14] predicted that nanopores of certain scale and structure on monolayer graphene would be very similar to biological ionic channels, but not yet confirmed. So we would prepare nanopores of a certain scale and structure on monolayer graphene by unfocused energetic $^{84}Kr^{25+}$ for investigating whether it would be similar to the ionic channels, and provide the selective channels for ions in the primitive ocean.

Moreover, a closure structure is very important and necessary for the graphene to localize concentration and provide space for biochemical reactions. Ugarte[15] reported that planar graphitic is easy to bend. Zhang et al[16] also reported self-assembly of graphene, and found that the graphene of a "closed" structure could be formed under some conditions. Ohtomo1 et al.[17] found proof of the existence of "onion" graphene (a hollow capsule of multilayer graphene) in Achaean Isua Metasedimentary rocks, which was considered as evidence of early life in the oceans at least 3.7 billion years ago. This indicated that the "onion" graphene with a hollow structure might be ubiquitously widespread in the primitive Earth. However, how did it roll up spontaneously to form a closure structure in the primitive ocean? In addition, phospholipid bilayer is also an important composition of cell membranes and the basis of cell growth and division[9,18]. Thus, how did it form on protocell membranes, and is this related to graphene? Furthermore, did graphene undergo some specific reaction with amino acids, especially with laevo-rotation amino acids (L-amino acids) and dextro-rotation amino acids? These resolutions would bring us into a graphene world with regard to the origin of life, and provide a possible route for designing a cell model and copying the process of the origin of life in laboratory.

## Results and Discussion

### GOs rolled up into capsules spontaneously and depended strongly on charge density and temperature

GOs (Fig. 1A) consisted of $sp^2$ carbons and $sp^3$ carbons. The $sp^2$ structure of graphite could be altered into $sp^3$ carbons by long-time oxidation (Fig. S2), causing visible roughness (Fig. S1)[19]. Then, high oxygen GOs (HOGOs, Figs. S2D-H) was obtained, and more cations were absorbed, which could enhance the protonation on the surface of GOs (Fig. S2C), finally improving the number of surface charges. High charge density would affect the surface tension of GOs, further bending the shape of GOs[20]. Thereby, as shown in Fig. 1B, contractive HOGOs could be observed in solution of low pH. Flocculation precipitations of HOGOs could be also found on the surface of saturated salt solutions of GOs at 4 $^{o}$C (Fig. 1G), suggesting that precipitation contains a mass of air[21]. $Pb^{2+}$, $Zn^{2+}$, and $Co^{2+}$ with higher charge for their impact of surface charge on bending of GOs could make GOs roll up in their saturated solution (Figs. 1C and S6). Although a few monolayer CGOs could be observed (Fig. 1C), most were multilayer CGOs (Fig. 1D) by TEM. These results showed that the formation of CGOs depended strongly on charge density and temperature, and the optimal temperature is 4 $^{o}$C (Fig. S4).

Moreover, due to thermodynamic instability, theoretical perfectly smooth surfaces do not exist on graphene. Ugarte[15] obtained curling and closure graphitic networks under electron-beam irradiation. Zhang et al. have proposed a mechanism in which graphitic sheets bend in an attempt to eliminate the highly energetic dangling bonds present at the edge of the growing structure[22]. Therefore, GOs with high charge density contract or bend in order to decrease the surface energy at low temperature, further decreasing the surface tension of GOs in solution (Fig. 2F). Then, the zero-dimensional materials, i.e., spherical CGOs, could be formed spontaneously to decrease surface energy. Fig. 1H shows the XRD of CGO compared with GO, reduced GO (RGO), and $C_{60}$. The 001 peak in $2\theta$ = 10.66$^{o}$ disappeared, but the 002 peak in $2\theta$ = 23.01$^{o}$ increased in RGO after reduction by Zn, indicating that the interlayer spacing of RGO decreased. The new peak of CGO appeared in $2\theta = 46.2^{o}$ after GO rolled up, while the 001 peak disappeared and the diffraction peaks of $Pb(NO_3)_2$ could be observed in CGOs, showing that the structure of CGOs changed into a unique structure between $C_{60}$ and GOs. However, the high concentration of $H^+$, $K^+$, and $Na^+$ could not make GOs roll into spheres or bubbles (Fig. S6), which may be ascribed to lesser charge than that of divalent ions. It is important to note that CGOs have not yet been observed in a non-solution system, perhaps as a result of the difference in preparation methods and systems.

**Sealed and stable capsules were formed by reduction**

Interestingly, a large number of crystals were found in some wrapped monolayer capsules under TEM (Fig. 2). As shown in Fig. 2B, the spacing of lattice fringe is 0.36 nm (Fig. 2E) and 0.73 nm (Fig. 2C) for coated layer and black crystals, respectively, indicating that the coated layer is monolayer graphene and the black crystal is $Pb(NO_3)_2$. These precipitated crystals should be come from the saturated salt solution during the formation of CGOs. The irradiation resembles a high temperature regime[15], so the salt crystals in CGOs were gradually melted and integrated together under the electron beam of TEM (Fig. 2A); however, the edge of monolayer CGOs appears quite regular and maintains a spherical or elliptical shape because of the presence of gas in closed capsules (Fig. 2B). The gas might be released from hydrochloric acid with the increase of temperature during preparation of TEM. Thus, the CGOs are unstable, and would unfold with the increase of temperature or decrease of salt concentration. Once the liquid spills from the unsealed edges, CGOs would then immediately return to the previous graphene oxide sheets. However, the electron beam from the TEM could not damage the whole structure of CGO, and the gas has always been enclosed in CGOs. This indicates that the CGOs have relatively strong stability of radiation, although it could be damaged by the radiation of an electron beam[23]. This is because divalent metals could possess coordination ability, which plays a role in the adhesive to “stitch” the opening of CGOs; this is a key factor in why the CGOs could not be observed in saturated solution of $K^+$ and $Na^+$ (Fig. S6).

To solve the problem of instability of CGOs, zinc powder was used to reduce CGOs at 4 $^{o}$C after sonication for 5 min, according to Mei and Ouyang[24]. Then some stable graphene capsules and a few nanotubes were obtained (Figs. 1E and 1F), after washing off the zinc powder using hydrochloric acid at room temperature. Almost all reduced CGOs (RCGOs) observed were multilayer, which may be due to the monolayer CGOs being quite fragile and hardly retained. Raman spectroscopy of RCGO shows a new peak of 1480 cm$^{-}$ between the G band and D band (Fig. 1J), which belong to *Ag(2)* band of fullerenes[25]. *Ag(2)* band is the contraction mode of five-membered rings, which indicates a sealed structure. So, the appearance of *Ag(2)* on RCGOs indicates that the sealed capsule of graphene is created by chemical reduction. Moreover, the peak of RCGO was lowered and shifted to 2930 cm$^{-}$ compared with the graphite peak of 2949 cm$^{-}$ on GO, affected by the chemical reduction of Zn (Fig. 1J). Fig.1I shows the XRD spectra of RGO, RCGO, and $C_{60}$. The 001 peak of GO in $2\theta$ = 10.66$^{o}$ disappeared in RCGO, a lighter and wider 002 peak of RCG in $2\theta$ = 23.01$^{o}$ retained in RCGO, indicating that the interlayer spacing of RCGO decreased after reduction by Zn. Meanwhile, the peaks of Zn and $Zn_5(OH)_8Cl_2 \cdot H_2O$ (simonkolleite) could be observed in RCGO. The simonkolleites might be reacted from the preparation course. Moreover, the existence of Zn showed that it was sealed into the RCGO, so it would not be washed off by concentrated hydrochloric acid and ultrasonic for three times. This also means that the RCGO is closer to an onion fullerene with sealed space, confirmed by the Raman spectrum above (Fig. 1J).

Surprisingly, the edge, shape, and size of monolayer CGOs (Fig. 2) seems to a cell removing layers of phospholipid molecules (Fig. 2D)[26]. So, could the CGOs' combined phospholipid be related to the origin of protocell membrane?

**GOs coated by phospholipid could also form a capsule structure**

The π rims on graphene serve as hydrophobic sites, and thus, the phospholipid hydrophobic end could be absorbed onto π rims on both sides of graphene[27]. Could the phospholipid bilayer structure also be formed on capsules of graphene oxide? UV spectrum shows that characteristic absorption peaks of GOs (~260 nm) are weakened by 1, 2-Dioleoyl-sn-glycero-3-phosphocholine (DOPC), indicating that the DOPC has successfully been coated on the surface of GOs; thus, the brownish red solution of GOs changes into a milk white color (Figs. 3A and 3B). The TEMs of GO and GO coated by DOPC (GO-DOPC) also show that many fine substances have been attached to the surface of GOs (Figs. 3A and 3B), causing the darker area with rough surface of GO-DOPC (Fig. 3B). Figs. 3E-G show AFM imaging in air, and the rough surface of GOs resulted in the fluctuant line of height; so, the minimum thickness was collected to compare GO and GO-DOPC. The thickness of GO and GO-DOPC is about 1.001 nm and 2.446 nm, respectively. It was reported that the thickness of a single layer GO is about 1.000 nm[28], indicating that the increased thickness (~1.445 nm) belongs to two layers of DOPC which coated on both sides of the graphene. Raman spectra (Fig. 3D) also confirmed that the three characteristic absorption peaks of GOs (band of 1D is ~ 1354, G is ~ 1601, 2D is 2939 $cm^{-}$, 1D/G = 1.41) have displaced after coating with DOPC (band of 1D is ~ 1349, G is ~ 1609, 2D is 2949 $cm^{-}$, 1D/G =2.14 ). Moreover, the 1D to G intensity ratio increased, which should be due to electron transfer from the *n*-doping DOPC in the lipid membrane[29]. Figs. 3G and 3H also indicate that the GOs coated by DOPC could also roll up. The layer fringes could be clearly observed by TEM in multilayer reduced CGOs with DOPC. The interlayer distance of DOPC-coated RCGOs (~0.55 nm, Fig. 3H) is bigger than that of multilayer RCGOs (~0.38 nm, Fig. 1D), indicating that there is bigger spacing in the interlayers of CGOs. As can be seen from AFM, CGOs exhibit an elliptical capsule according to length and height (Fig. 3G). Fig. 3K shows the TEM of monolayer CGO-DOPC, and the surface of the capsules is stained black burr-like substances. The metal ions could be adsorbed on P or N of DOPCs, forming the burr-like composites of Pb and DOPC. The hydrophobic groups of DOPC in these composites could be adsorbed on the π rim of GOs; thus, a large number of burr-like composites would be observed on the surface of CGO. Moreover, Fig. 3M shows an enlarged image of Fig. 3K, in which the bilayer structure could be clearly observed and closely resembles the phospholipid bilayers of cells (Figs. 3I and L) [30].

Some researchers proposed that fatty acids may be candidate materials of protocell membranes[9,31,32]. However, fatty acid membranes have two defects: 1) their lack of selective channels, such as ion channels; and 2) their lack of stable structure support. Thus, they hardly survived in the harsh environments of the early Earth. Fortunately, graphene could compensate for these shortcomings. After the formation of GOs with phospholipids and protein analogs, graphene would be damaged by the reactions between compound groups with continuous evolution, and then be gradually cut[33] and digested, finally evolving into a network of aromatic amino acids. These events were confirmed by Doyle through the structure of $K^+$ channels[34] (Fig. S15). Interestingly, Tu et al. reported that the phospholipids of cells were penetrated and extracted by graphene nanosheets, leaving a cell without a phospholipid frame, but appearing stable and unbroken[26], indicating that some substance remains to support the cell structure, except the phospholipid bilayer; this could perhaps be the network of aromatic structure originating from graphene. Consequently, the CGOs coated with phospholipid bilayer could be proposed as the candidate embryo of membrane frame for protocells.

**GOs represent more affinity for L-amino acids, and may be the original drive to promote the formation of left-handed bio-based protein**

Proteins are one of the most important constituents for cell membranes. For this reason, the adsorption of base unit amino acid chains on the surface of GOs has been studied in detail[35]. However, the absorptive capacity of amino acids of different optical activity was not performed; thus, we compared the binding ability for L-amino acids and D-amino acids on the surface of GOs under certain conditions. The effect of pH and temperature on the *N/C* of products between GOs and L-proline was firstly studied to optimize the experimental conditions (Figs. 4A and 4B). pH 3 and 90 $^{o}$C were selected for this purpose. Figs. 4C and 4D show the XRD and infrared spectroscopy of reaction products between GOs and amino acids. *C-N* bonds (~1380 $cm^{-}$) were observed in Fig. 4D, demonstrating that the composites between amino acids and GOs have been synthesized at 90 $^{o}$C. Meanwhile, although the 001 peak of GOs has disappeared after reaction, the 002 peak of graphene in $2\theta = 23.01^{o}$ could be found in all samples of GO-amino acids (Fig. 4C), indicating that the spacing distance of GO- amino acids has decreased. Meanwhile, a lighter peak at about $2\theta = 45^{o}$ could be observed in GO-L-proline, meaning that the interlayer spacing between GO and L-proline composite is smaller, which is due to the stronger affinity between GO and L-proline. Consequently, the *N/C* of reaction products between amino acids and GOs were higher in GO-L-amino acids than that of D-amino acids and GOs. The *N/C* and $\Delta N/C$ are proportional to the weight of amino acids, as shown in GO-arginine, proline and glycine (Table 1), which is ascribed to the activity of amino acids. These could be further confirmed by thermogravimetric analysis (TGA), as shown in Fig. S12, the mechanism of which is discussed in detail in the Supplementary Information. These results indicate that amino acids could easily react with graphene, and that L-amino acids easily form peptide bonds. Katoch et al.[35] reported that the peptides adsorbed on graphene could change into secondary structures because of their interaction with the surface of GO, implying that these peptides connected at graphene might slowly evolve into membrane proteins, agreeing with the theoretical predictions of Zhan[36]. Moreover, the difference in the reaction on GOs between laevo-rotation and dextro-rotation amino acids might be related to the left-handed bio-based amino acids of modern life. If the CGOs are the candidate for the protocell membrane frame, then we need to consider whether graphene could evolve into some functions similar to cell membranes, such as ion selectivity.

**Nanopores of high ion selectivity were obtained on graphene**

The GOs membrane with defects has a certain degree of selectivity for ions and gas molecules[37-39]. However, since it is very difficult to prepare a membrane of monolayer GOs to separate ions, we have to use graphene instead of GOs in order to obtain controllable nanopores. Therefore, a nanopore graphene membrane with polyethylene terephthalate (MNGM-PET) was prepared by irradiating energetic ions of $^{84}Kr^{25+}$ and chemical etching according to the method of Russo et al.[40] and O'Hern,

S. C. et al.[41], the processes of which is shown in Fig. 5A. The results demonstrated that the small and large pores of tapered holes on both sides of the single PET membrane were 10–20 and 50–80 nm, respectively, (Figs. 5B and 5C). The nanopore size on the PET side of MNGM-PET was similar to the nanopores on a single PET membrane (40 nm to 60 nm, Figs. 5B and 5D). The pores on the graphene side of MNGM-PET were too small to be observed using a scanning electron microscope (SEM) or transmission electron microscope (TEM) even with a resolution ratio of 1 μm to 2 μm (Fig. 5E). A higher resolution ratio (1 μm to 2 μm) could not be used for graphene as it would be damaged by the electron beam necessary to achieve a higher resolution ratio. According to Russo et al., who also obtained nanopores produced through the irradiation of unfocused energetic ions ($Ar^+$), with radii as small as 3 Å (correspond to 10 removed ) based on the azimuthal integral, our nanopores should fall in this scale range (< 5Å)[40]. Then, MNGM-PET was contained between the sources and driving sinks to study the ion permeation (Fig. S11). Considering that each ion needs a long time (over weeks) to completely permeate into the driving sink, ions would be inputted into the source sink to find the priority ion crossing the MNGM-PET, and then this ion would be deleted to find the next priority ion. The results have shown that monovalent metal ions could rapidly cross MNGM-PET, and the permeation sequence is high ion selectivity as, $Rb^+ > K^+ > Cs^+ > Na^+ > Li^+$ (Fig. 6), which is substantially similar to the selectivity sequence of $K^+$ channels in the cell membrane ($Rb^+$, $K^+ > Cs^+ > Na^+$, $Li^+$)[34].

**Ion selective mechanism of nanopores on graphene**

Doyle et al.[34] explained that the $K^+$ selectivity permeation is influenced by carbonyl oxygen atoms, which provide multiple closely-spaced sites. So, the permeation is constrained in an optimally fit ion geometry, resulting in a dehydrated $K^+$ with proper coordination to cross, but which is extremely small to dehydrated $Na^+$. In the present study, the nanopore scale could not be controlled through high energetic ion irradiation. According to Doyle et al., nanopores with optimally constrained geometry fits for every dehydrated ion are available. Therefore, each ion would pass through the optimally fit pore on MNGM-PET with similar speeds, and competition would not occur among ions. However, high ion selectivity was observed, and competition occurred between ions, such as $Rb^+$ inhibited $K^+$, as shown in Figs. 6A and 6C. Russo et al.[40] confirmed that the nanopores produced through the irradiation of energetic ions achieve the normal distribution with differential scales. However, the "filter tip" of $K^+$ channels is 10 Å[34]. Therefore, the fit optimal geometry could not exist on MNGM-PET, and Doyle's theory could not be used to explain the ion selectivity of MNGM-PET. Celebi et al.[38] have reported that smaller nanopores on graphene possess higher separation ability. We considered that as long as the nanopore scale is adequately small, then the ion selectivity conferred by these nanopores would be able to separate every ion. Thus, this high similarity between MNGM-PET and $K^+$ channels for ion selectivity might imply a common characteristic. In fact, the $K^+$ channels in the cell membrane also have a similar structure to MNGM-PET, and consist of a small "filter-tip" with carboxyl groups located in the middle of the channel; the two ends of the "filter-tip" are narrow conical tubes composed of proteins[34]. Aside from the permeation sequence similar to a $K^+$ channel, the high permeation speed needs to be also considered.

Ion permeation across MNGM-PET is rapid before 30 s (Figs. 6, 7, and S13), and reaches ~3.5 mg/L (~$3.56 \times 10^{15}$ ions per second) for $K^+$ (Fig. 6), which is significantly higher than the cell membrane ($10^8$ ions per second)[34]. The ion permeation also exhibits a throughput rate exceeding the diffusion limit, implying strong energetic interactions between metal ions and the pore[34]. To explain the 0-30 s stage, we need to consider the MNGM-PET structure and composition. When ions with high charges and energies run through MNGM-PET, atoms on their paths would be ionized and carried with radicals[42], and easily oxidized into carboxyl or hydroxyl groups[41]. Thus, nanopores could contain carboxyl groups on MNGM-PET in this study. Oxygen-containing groups with negative charges on the MNGM-PET nanopores could strongly attract cations, and thus, metal ions with high charges would be more easily attracted and primarily permeate through the nanopores. Furthermore, the permeation sequence strongly depended on the velocity within the electric field[39]. As hydrated ions move at high speeds, water molecules will then be stripped. This means that the ions passing through MNGM-PET nanopores would be bare, and the sequence would not depend on the hydrated ionic radius. Based on this consideration, the permeation sequence would only depend on the metal ion mobility (Table S1). The sequence would be $Cs^+$ > $Rb^+$ > $K^+$ > $Na^+$ > $Li^+$, which is consistent with experimental results, except for $Cs^+$ ($Rb^+$ > $K^+$ > $Cs^+$ > $Na^+$ > $Li^+$). $Cs^+$ is a weak Lewis acid with a low charge-to-radius ratio that weakly interacts with ligands[43]; thus, the special properties would make $Cs^+$ follow $K^+$. Other studies also reported that $K^+$ channels exhibit a selectivity sequence of $K^+ \approx Rb^+ > Cs^+ > Na^+ \approx Li^+$ [34,44]. Moreover, this sequence could also explain the origin of the cell selectivity between $Na^+$ and $K^+$, in which almost all cells preferentially absorbed $K^+$, but not $Na^+$. $K^+$ is at least 10,000 times more permeable than $Na^+$[34], even though the latter is the most abundant ion in seawater[45], and its radius and chemical properties are extremely similar to $K^+$[34]. Moreover, the pattern of rapid filtration before 30 s can be described as the Poisewille formula [46] for liquid laminar flow (seen in the Supplementary Information). The calculated rate of $K^+$ (~9.35 mg/L) is close to the measured value (~8.75 mg/L, Fig. S14).

From 30 s to 16 h (Fig. 6), the slow permeation rate could be attributed to the carboxyl ion exchange in the nanopores [47]. Carboxyl groups on the small nanopore of graphene function as “filter tip” and could provide a medium for ion exchange between permeation and the driving liquid, selectively allowing metal ions through MNGM-PET. When metal ions selectively cross the MNGM-PET nanopores into the driving liquid, $H^+$ moves across the PET membrane nanopores along with the pH gradient, and a high pH gradient induces the high ion permeation (Fig. S14), indicating more driving force. As the primitive ocean is strongly acidic and has high salt content [7], the high ion-selective and driving effect of $H^+$ is slightly comparable with the function of adenosine triphosphate in modern cells. Moreover, the rate of permeation is substantially lower than the initial stage. However, $Cs^+$ is an odd ion in this stage because its content decreases in absence of $Rb^+$ and $K^+$, and $Na^+$ content increases conversely. The decreasing $Cs^+$ content could be attributed to this ion's adsorption onto the inner surface of the nanoholes in the MNGM-PET PET basal layer. $Na^+$ could probably improve $Cs^+$ adsorption onto groups with negative charges, and thus, a peak valley could be observed in the absence of $Na^+$ and presence of $Li^+$ (Figs. 6C-G), indicating that $Cs^+$ can rapidly reach the minimum adsorption and slowly increase again through ion exchange.

In addition, the peak of $Cl^-$ could not be observed in all driving liquids of PET and MNGM-PET (Figs. 6I and 6J), and the flocculent precipitate of AgCl could not be seen in samples, suggesting that the $Cl^-$ could not be permeated through MNGM-PET and PET from source liquid to driving liquid. This is due to the chemical structure of nanopore in PET and MNGM-PET, in which the oxygen groups in the wall of nanopore could prevent the permeation of $Cl^-$.

**The nucleotides might originate from graphene tubes**

Although many researchers have asserted that the formation of nucleotides might need a suitable synthetic template, it has not yet been found or confirmed. Zhan[36,48] predicted by calculation that the spiral angle ($30^o$ and $60^o$) of carbon nanotubes are conducive to the formation of DNA molecules (spiral angles of *A-DNA* and *B-DNA* are $30^o$). Moreover, the height of the repetition period is $d_{h\text{-}C}$ = 0.142 nm in the vertical direction, and the distance of the spiral period is $d_{S\text{-}C}$ = 0.426 nm for nature carbon nanotubes. The length of the repetition period of the spiral direction is $d_{S\text{-}D}$ = 0.64 nm for A-RNA, and the distance between the two phosphorus atoms of the repetition units is $d_{h\text{-}P\text{-}A}$ = 0.59 nm and $d_{h\text{-}P\text{-}B}$ = 0.71 nm for A-DNA and B-DNA, respectively. So, the following relationship can be calculated: $2d_{S\text{-}D} \approx 3d_{S\text{-}C}$, $2\ d_{h\text{-}P} \approx 5\ d_{h\text{-}C}$, $3\ d_{h\text{-}P} \approx 5\ d_{S\text{-}C}$. These relations show that the structures of A-DNA and B-DNA fit the carbon nanotubes of (10, 0) and (5, 5), respectively (Fig.7A). Katoch et al.[35] reported that the peptide forms a complex reticular structure upon adsorption on graphene and graphite, and the dominant structure in the powder form is α-helix, which is consistent with predictions of Zhan that α-helix proteins would be formed in carbon nanotube. Guerret et al. synthesized a spiral of metal nanowires through carbon nanotube[49]. These (Fig. 7A) imply that graphene tube might provide a template for the formation of nucleic acid molecules in the early Earth. A few graphene tubes were obtained and mixed in RCGOs (Fig. 1F). It was also fabricated experimentally by Zeng et al. (55), but no CGOs were obtained; this may be due to the difference in preparation methods. So, if the graphene tubes exist in primitive Earth, it could be used as the template of DNA in some conditions.

**A possible evolution process of protocell membrane on graphene embryos**

A possible evolutionary process of life involving embryos of graphene could be presented as follows (Fig. 7B). The GOs, originated from the primitive atmosphere in hypoxic conditions under the effect of lightning, could roll into CGOs in the primitive ocean. A part or whole CGOs could be repeatedly opened or rolled up by changing the temperature and salinity, and graphene tube might provide a template for the formation of nucleic acid molecules. Then, numerous minute molecular organic compounds, such as amino acids and nucleotides, would be adsorbed and then wrapped into the inside of CGOs, resembling original organic substance exchange. Furthermore, the ubiquitous cosmic ray, primarily composed of high-energy protons and atomic nuclei with positive charges[50], could defect GOs to produce nanopores on their surfaces, similar to ion channels. With the change of the geological environment of the primitive ocean, the selectivity which removes high content of negative $Na^+$ in CGOs, determines the amount and type of ions inside CGOs, such as $K^+$ in cells. Meanwhile, as the double surfaces of a GO monolayer contain abundant carboxyl groups and π rims, peptide comprised mainly of L-amino acids and phospholipid analogues would be absorbed[51] and would have gradually evolved into primitive cell membranes with bilayer phospholipids and proteins. The graphene embryos would have gradually been cut[33], digested, and evolved into a network of aromatic amino acids[34] through the reactions between functional groups during evolution. The protocell membrane would then be formed with liquidity for division, stability for protection, and high ion selectively essential for life activities, which created and expanded life gradually.

## Materials and Methods

### Formation of CGOs

GOs with differential content of oxygen were prepared by controlling oxidation time (8 h, 18 h, 30 h, 44 h, and 70 h) using improved Hummers method. These materials were characterized by XPS, Raman, FTIR, and Automatic Potentiometric Titrator so that the oxygen level of samples could be determined. The GOs with oxidation time for 70 h were called HOGOs. About 4 mg HOGOs was added into 2 mL of HCl (pH 0.63). Then, saturated solutions of $NaNO_3$, $KNO_3$, $Pb(NO_3)_2$, and $Zn(NO_3)_2$ at 4 ℃ after strong shocking, respectively, were collected for preparing samples of TEM and AFM. 4 mg HOGOs and 20 mg zinc powder were then added into the 2 mL saturated solution of $Zn(NO_3)_2$ at 4 ℃ and sonication was performed for 5 min in order to reduce CGO (RCGO)[24]. 4 mg HOGO was added into saturated solution of $Pb(NO_3)_2$ at 4 ℃, and acidity was adjusted to pH of 0.15. After strong shocking, it was collected for preparing samples of TEM and AFM. The details of the optimization experiments can be found in the External Databases.

### Preparation of CGO coated by DOPC

20 mg of DOPC was added into the 5 g/L suspension of HOGOs *via* ultrasonic treatment for 1 h, and then centrifugal washing with water and ethyl alcohol was used to obtain the HOGOs coated by DOPC (GO-DOPC). It was then collected for characterization with TEM, AFM, UV, and Raman. After drying at 40 $^{o}$C for 24 h, 10 mg of GO-DOPC was added into saturated solution of $Pb(NO_3)_2$ at 4 $^{o}$C. Strong shocking and centrifugal washing at 4 $^{o}$C were performed, and then the samples were collected for characterization with TEM and AFM.

### Affinity of amino acids for GOs

About 22 g of $Pb(NO_3)_2$ and 10 mg of GOs were added into 5 mL of deionized water and ultrasonicated for 10 min. Afterwards, 1.5 mL of glycine, L,D-proline, and arginine were added into the above mixture and ultrasonicated for 10 min, respectively. The above mixture was then placed in a silicone oil bath, and stirred for 16 h at 50, 70, 90, 110 and 160 ℃ in pH of 1, 3, 5, 9 and 12, respectively. The products of 90 $^{o}C$ and pH 3.5 were repeatedly washed with $H_2O$ and acetone, and were analyzed using XRD, FTIR, elemental analysis, and thermogravimetry analysis after drying for 24 h.

**Preparation of MNGM-PET**

The graphene monolayer with copper basal membrane, purchased from XFNANO Materials Tech Co., Ltd. (Nanjing, China), was transferred to polyethylene terephthalate (PET) membrane and irradiated with 1915 MeV $^{84}Kr^{+}$ ions in a heavy ion accelerator at the Institute of Modern Physics, CAS. The sample materials were imbedded between two etching baths, and 0.5 M HCl was poured into one bath close to the graphene, while 9 M NaOH was poured into the other bath close to PET (Fig. 1a). Two platinum electrodes supplied by a source meter (Keithley 6487) were inserted into the left and right baths, and a 0.1 V controlled voltage was applied to initiate the etching. The electric current was suddenly increased to remove the etchant and stop the etching. The samples were repeatedly washed until neutral. Finally, MNGM-PET was removed for shape, size, and structure characterization through SEM. The single PET with conical pores was prepared according to the same procedure used for the control group.

**Permeation of ions**

Primitive seawater is strongly acidic (pH ≈ 1.25, 3.8 billion years ago[45]), and PET is unstable after heavy irradiation in basic solution. Therefore, HCl was used as the driving liquid. MNGM-PET was installed in the permeation device, and HCl (pH 1) and solutions of monovalent metal salts, such as LiCl, NaCl, KCl, RbCl, and CsCl (0.1 mol/L per ion, pH 2.5) were injected into the driving and source sink, respectively. A certain volume of the driving liquid was collected at different times, and analyzed for metal ion content across MNGM-PET using ICP-AES (IRIS Advantage ER/S, TJA). As each ion required a long time to completely permeate into the driving liquid, we designed a selectivity experience in which ions would be inputted into the source sink to study which ion would firstly cross the MNGM-PET within 16 h. This ion would then be deleted, and the other ions would be inputted again into source sink to determine the second priority ion that would cross the MNGM-PET. The experiments were performed in cycles to determine the source sequence of the monovalent ions. The slope of the line represents the source ratio of the ions.

$HNO_3$ (pH 1) was then poured into the driving sink, while KCl (0.1 mol/L, pH 2.5) was poured into the source sink. Afterwards, a certain volume of sample was collected from the driving sink after 16 h, and the $Cl^-$ content of the sample was determined using ion chromatography (861 Advanced Compact IC, Switzerland). A small amount of $AgNO_3$ was added into the sample to determine whether AgCl had been generated.

## Acknowledgments

This study was conducted with financial support from the National Natural Science Foundation of China (No. J1210001, 31201841, and 21405165), and Western Light of CAS (Y212130XB0). The authors thank Prof. Tim J. Flowers and Ting Zhu for their thoughtful comments on this manuscript.

## Author contributions

Li Zhan, Yao Huijun, Wang Chunmei designed the project, while Li Zhan and Wang Chunmei wrote the main paper. Yao Huijun prepared and characterized MNGM-PET, while Li Zhan and Tian longlong conducted the permeation experiments and processed the data. Wu Wangsuo revised the paper, and all authors reviewed the paper.

**Competing financial interests:** The authors declare no competing financial interests

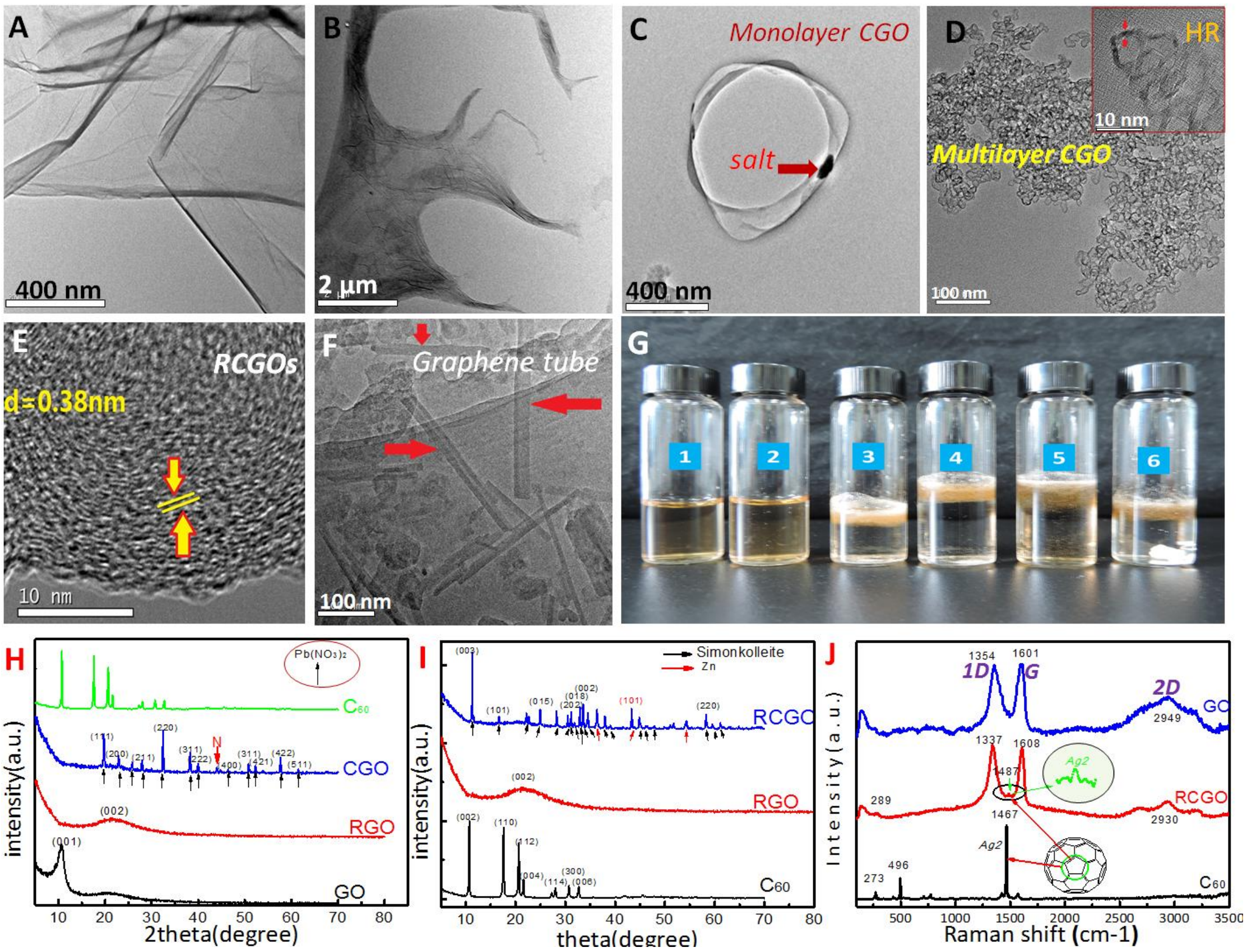

Figure. 1 The CGOs under different conditions. (**A**), the TEM of GOs. (**B**), the TEM of HOGOs at 4 $^{o}C$ in pH 0.62. (**C**), the TEM of monolayer CGOs in saturated solution of $Pb(NO_3)_2$ at 4 $^{o}C$ in pH 3.5, the red arrow points to a salt particle. (**D**), the TEM of multilayer CGOs in saturated solution of $Pb(NO_3)_2$ at 4 $^{o}C$ in pH 3.5, HR represents high-resolution of **D**. (**E**), the TEM of a reduced CGO(RCGO), the CGOs were reduced by zinc at 4 $^{o}C$ in pH 3.5 and sonication for 5 min, then washed with HCl at room temperature for removing the zinc, the spacing of lattice fringe is 0.36 nm. (**F**), the TEM of graphene tube roll, CGOs were reducted in the same condition as **E**, the arrow points to the tube roll of graphene, the hollow structure of tubes can be seen clearly. (**G**),the precipitation of GOs in different conditions, **1** is suspension of GOs, **2** is in HCl of pH 0.62, **3** is in saturation of NaCl, **4** is in saturation of NaCl at pH 0.15, **5** is in saturation of $Zn(NO_3)_2$, **6** is in saturation of $Pb(NO_3)_2$. All samples are at 4 $^{o}C$. (**H**), the XRD of CGO, $C_{60}$, RGO and GO, Pb represents $Pb(NO_3)_2$ ,red N represents a new peak. (**I**), the XRD spectrum of GO, RCGO and $C_{60}$, RCGO is the reducted CGOs by zinc in 4 $^{o}C$ under sonication for 5 min, simonkolleite is $Zn_5(OH)_8Cl_2$•$H_2O$.(**J**), the Raman spectrum of CGO of GO, RCGO and $C_{60}$, a new peak (1487 $cm^{-}$) of *Ag2* between 1D (1337 cm$^{-}$) and G brand (1608 cm$^{-}$) could be seen in RCGO. Each spectrum was the average of 1000 scans, at 2 cm$^{-1}$ nominal resolution.

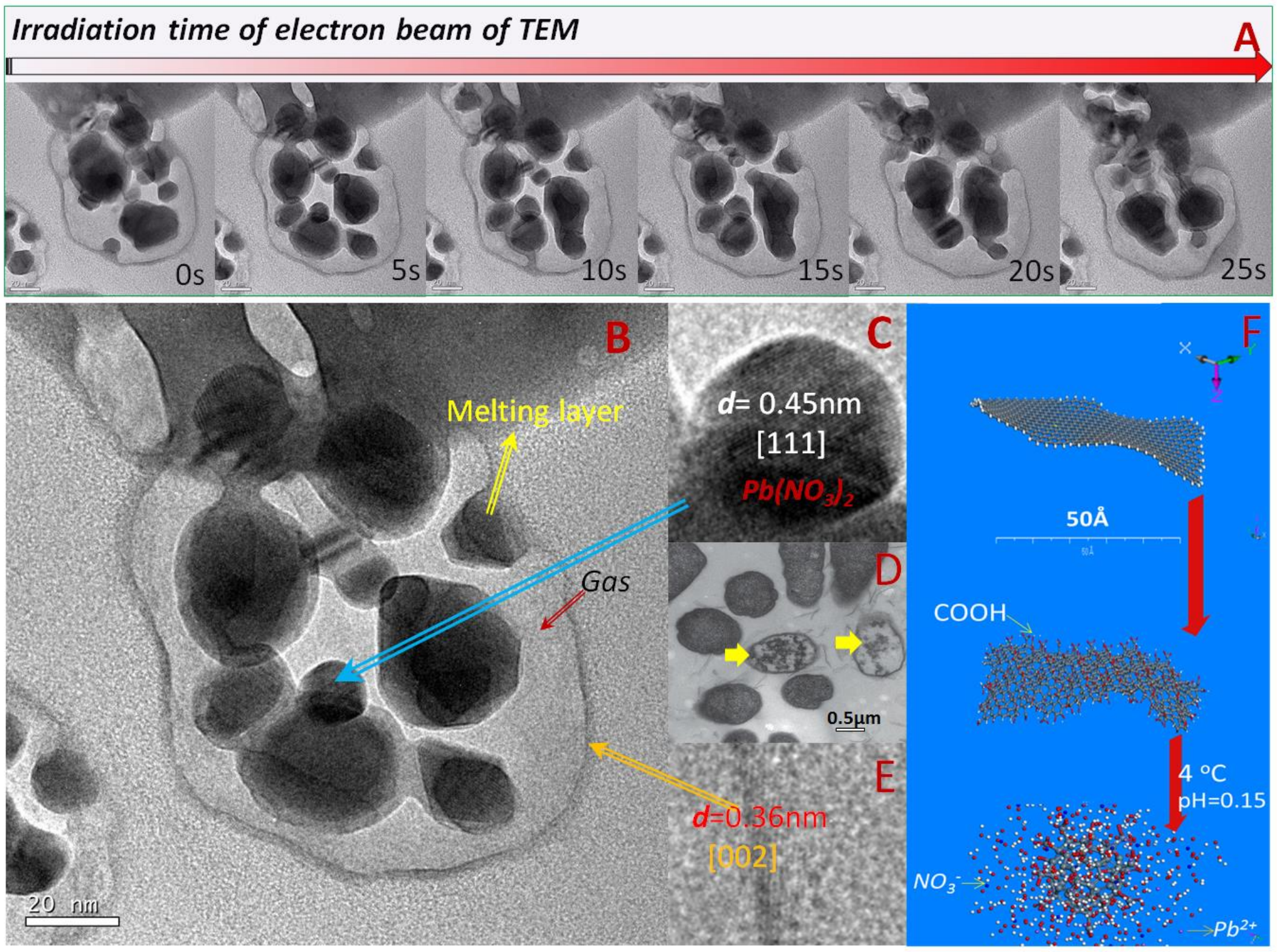

Figure 2. Comparison of the monolayer CGO and bacteria. (**A**), the dynamic map of irradiation on morphological changes of CGO in TEM, the CGO was prepared using HOGO at 4 °C and saturation solution of $Pb(NO_3)_2$, as well as pH 0.15, the long red arrow represents the direction of time. (**B**), crosssection of the monolayer CGOs through TEM, the melting layer is the melting of metal salts under electron beam, the fusion process of melting could be clearly seen from Fig. A. (**C**), a partial enlarged view of the salt crystals in the **B**, ***d*** is the spacing of the lattice fringe. 0.45 nm is the [111] crystal plane $Pb(NO_3)_2$. (**D**), the picture of bacteria without phospholipids from the paper of Tu et al. published in *Nature Nanotech* in 2013[26]. The yellow arrow points to the bacteria without phospholipids which was extracted by GOs. (**E**), the coating layer of graphene, the d=0.36 is the [002] crystal plane of graphene. **(F)**, the diagram of the graphene roll up.

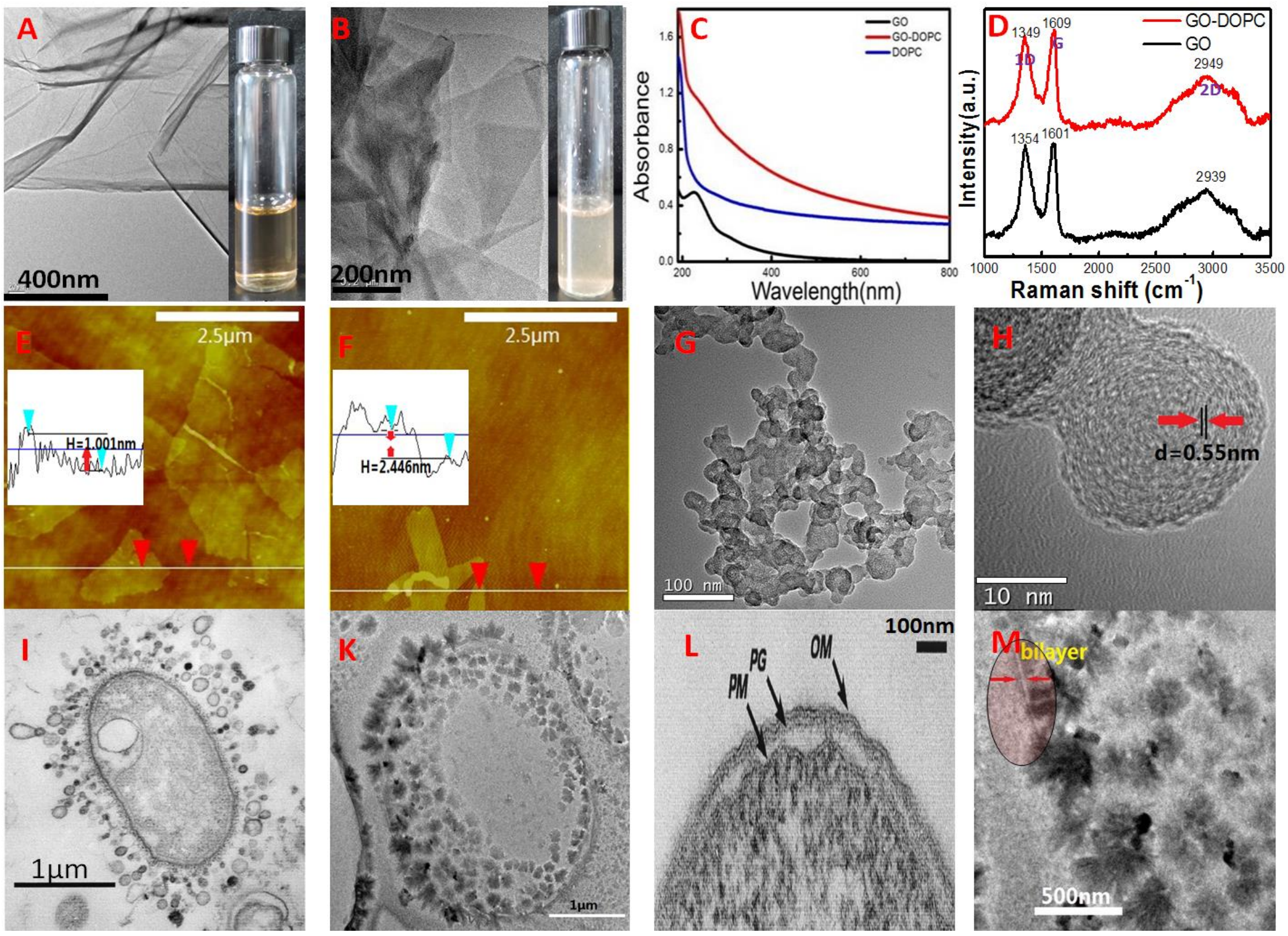

Figure 3. The CGOs with phospholipid bilayer. (**A**), the TEM of GOs, the solution of GOs is in a bottle. (**B**), the TEM of GO with phosphlipid bilayer (GO-DOPC), the solution of GOs-DOPC is in a bottle. (**C**), the UV Spectra of GOs, GOs-DOPC, and DOPC.; (**D**), the Raman Spectra of GOs, GO-DOPC. Each spectrum was the average of 1000 scans, at 2 $cm^{-1}$ nominal resolution. (**E-F**), the AFM of GOs and GOs-DOPC, the curve in the white pitures show the line profiles indicated in images of AFM, respectively, the *H* is the different height levels. **(G)**, a larger number of multilayer CGO-DOPC at 4 $^{o}C$ in pH=3.5 and saturation solution of $Pb(NO_3)_2$. (**H**), the TEM of a multilayer of CGO-DOPC after redution of *Zn* powder, ***d*** is the distance of layer fringes. (**I**), the TEM of an unidentified gram-negative bacterium from ref. (J. Bacteriol. 1999, 181(16):4725,[30]). (**K**), the TEM of monolayer CGO-DOPC, black burr-like substance is the composites of DOPC and metal ions, prepared at 4 $^{o}C$ in pH=0.15 and saturation $Pb(NO_3)_2$. **(L)**, thin section of the cell envelope of *E. coli* K-12 after conventional embedding. The periplasmic space is empty of substance, and the peptidoglycan layer (PG), outer membrane (OM), and plasma membrane (PM) can be seen from ref. (J. Bacteriol. 1999, 181(16):4725). (**M**), the enlarged TEM of **K**, the bilayer membrane of DOPC on GOs could be seen in the circle, the glitch directionality could show clearly the 3D structure of CGO-D.

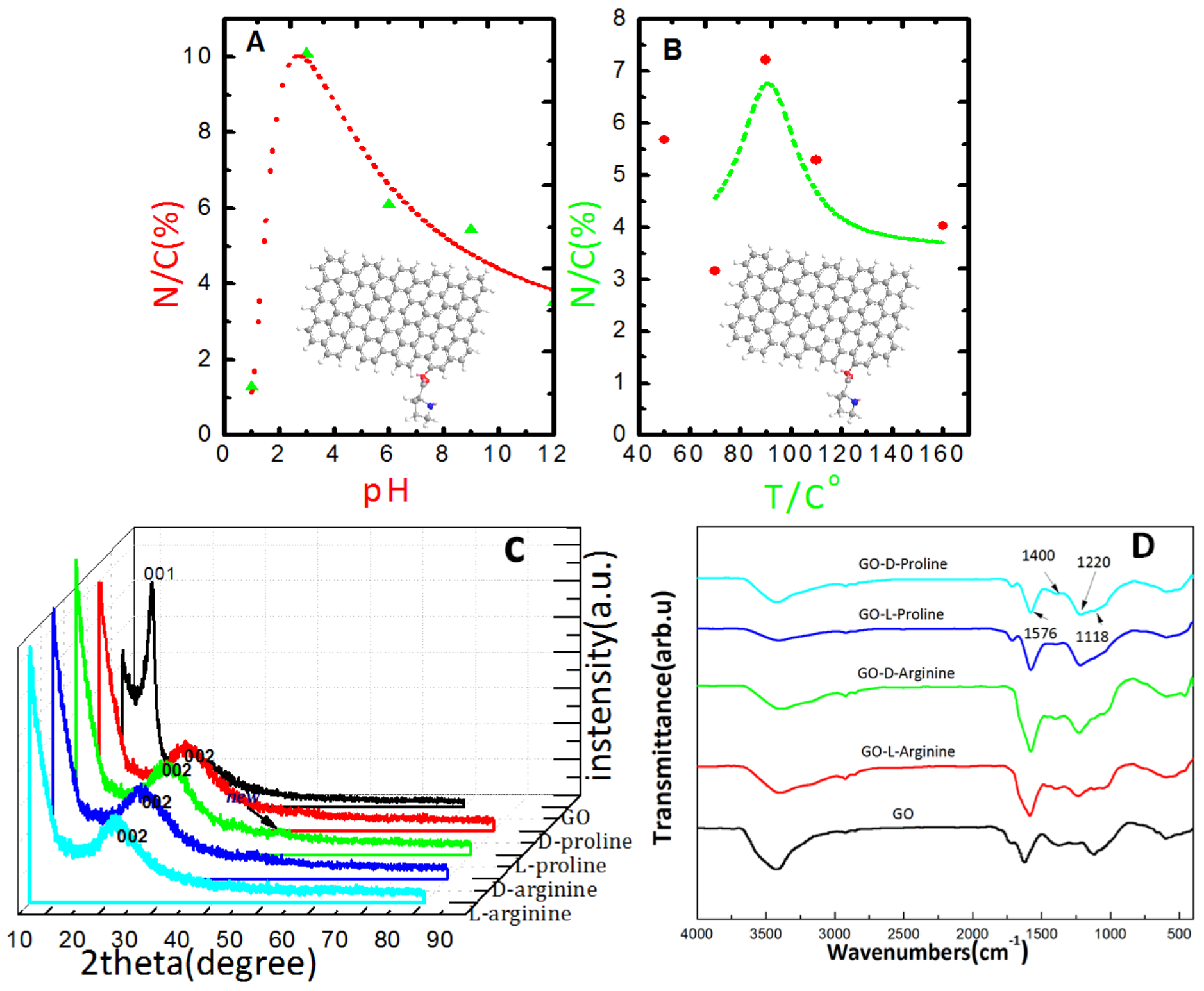

Figure 4. (**A**), the effect of pH on the *N/C* of reaction products between GOs and L-Proline, *N/C* was determined by elemental analyzer, pH is 1, 3, 5, 9 and 12. *N/C* is the ratio of *N* to *C* content in GO-amino acids. (**B**), the effect of temperature on the *N/C* of reaction between GOs and L-Proline, temperature is 50, 70, 90, 110 and 160 $^{o}$C. *N/C* is the ratio of *N* to *C* content in GO-amino acids. (**C**), the XRD of composite between amino acids and GOs, samples were prepared in pH of 3 at 90 $^{o}$C. (**D**), the FTIR Spectrometer of composite between amino acids and GOs, samples were prepared in pH 3 at 90 $^{o}$C.

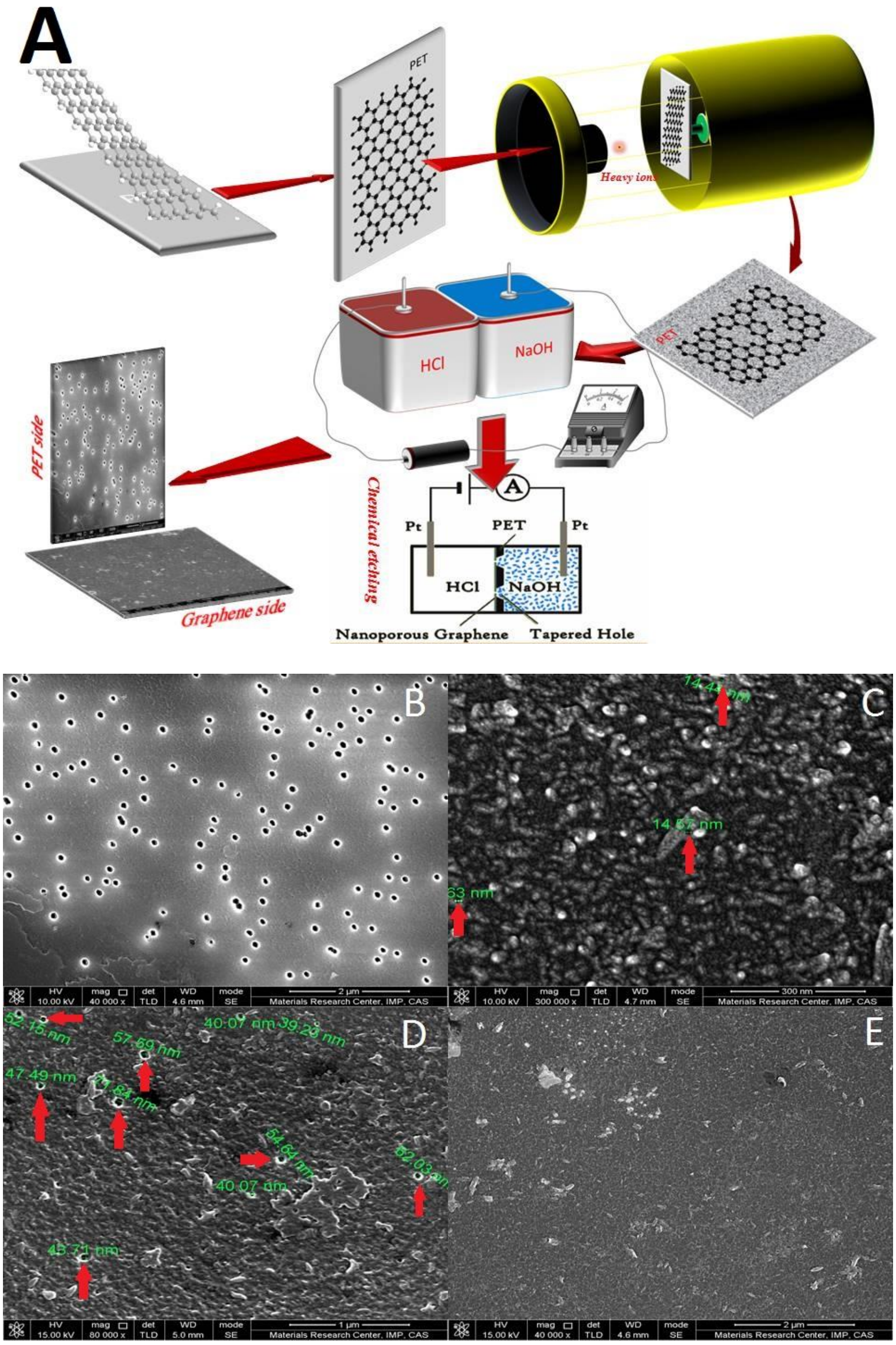

Figure 5. The preparetion processes of MNGM-PET **(A)** SEM of PET. **(B-E)** MNGM-PET. **(A),** The heavy ion is 1915 Mev $^{84}Kr^{25+}$, [HCl] = 0.5 M, [NaOH] = 9 M, *V* = 0.1 V, the thicker of PET bascal layer is about 20 µm. **(B),** the bigger scale of the hole for a single PET. **(C),** The smaller scale of the hole for single PET. **(D),** the bigger pore on PET basal layer side of MNGM-PET. **(E),** the graphene side of MNGM-PET.

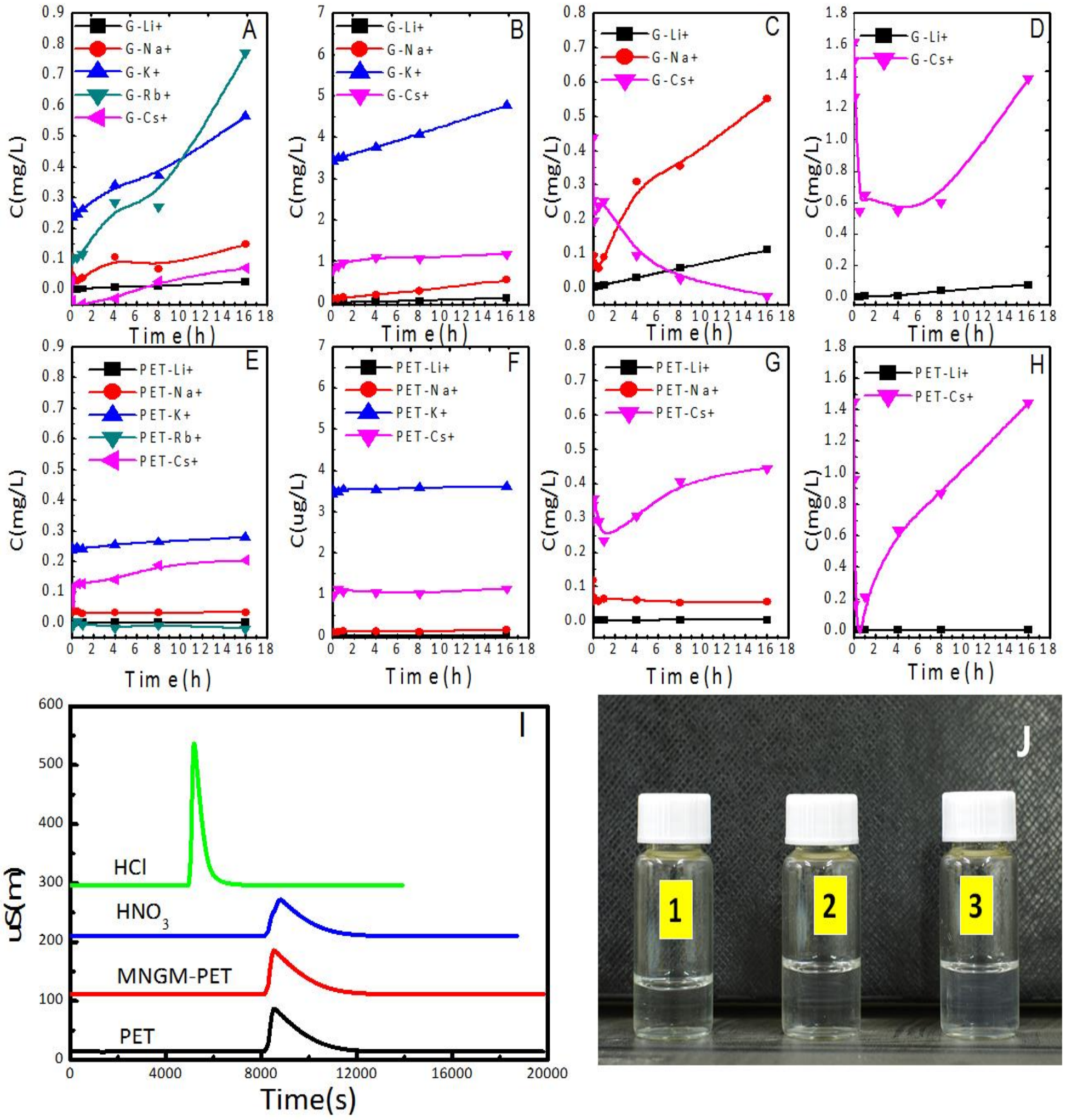

Figure 6. The separation of MNGM-PET and PET membrane for $Li^+$, $Na^+$, $K^+$, $Rb^+$, $Cs^+$, and $Cl^-$. ΔpH = 1.5, $t$ = 25 °C, $C_{ions}$ = 0.1 mol/L. (**A**), (**B**), (**C**), and (**D**) are the MNGM-PET filters. (**E**), (**F**), (**G**), and (**H**) are the PET filter. The Y-axis is the ion concentration that crossed the membrane. (**I-J**), $Cl^-$ filtration across the MNGM and PET membrane. ($\Delta pH$=1.5, $t$=25°C, $C_{ions}$=0.1mol/L). (**I**) is $Cl^-$ analysis using ion chromatography. The peak of $NO_3^-$ can only be observed in the sample of MGNM-PET and PET, the $Cl^-$ could not be found in all samples. (**J**) is $Cl^-$ concentration analysis with $AgNO_3$. 1 was the control group, 2 was the MNGM-PET 3-h after filtration, 3 was the PET membrane 3-h after filtration, AgCl precipitations could not be observed in all samples

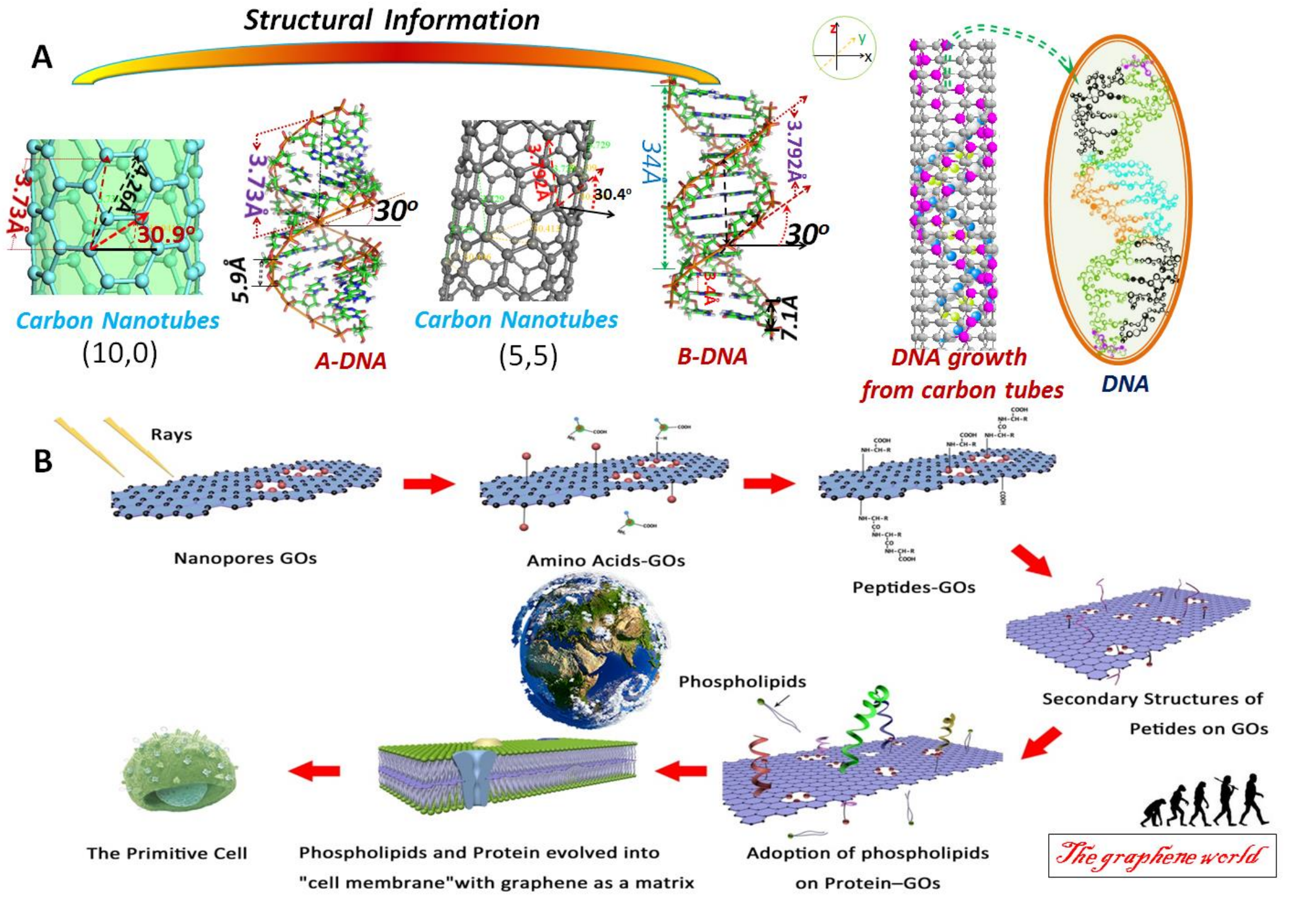

Figure 7. (**A**), structural comparison between graphene and DNA, and the schematic diagram of the origin of DNA from carbon nanotube. (**B**), the evolution scheme of cells with graphene as an embryo, the Figure 7B was drawn by the one of co-authors using Chemical Office 2010; the part of idea referenced from paper of *Zhan [36]*.The GOs and phospholipids can be found near/in volcanos under the sea or the primitive atmosphere. Thus, we assumed that all processes occurred in the seawater at the early age of Earth. The rays were mainly cosmic rays

Table 1. The elemental contents of *C* and *N* in composites of GO and anmio acids

| *Groups* | *Amino acids* | $N/C_{L\ or\ D}$ | $\Delta N/C$ |
|---|---|---|---|
| *GO-Arginine* | 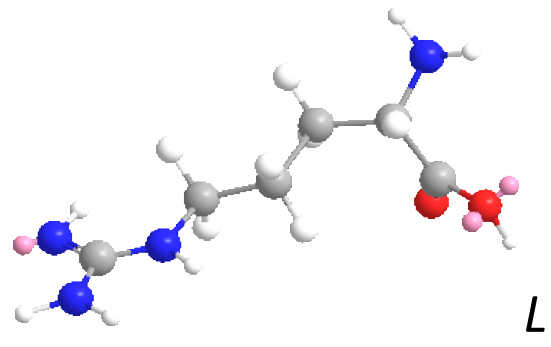 *L* | 19.86±0.13% | 18.36% |
| | 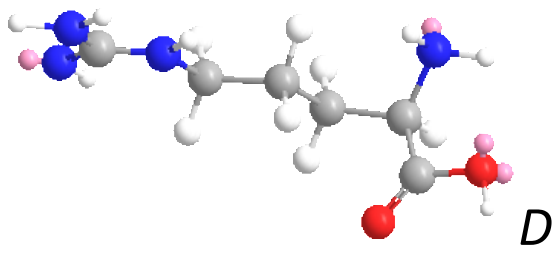 *D* | 16.78±0.08% | |
| *GO- Proline* | 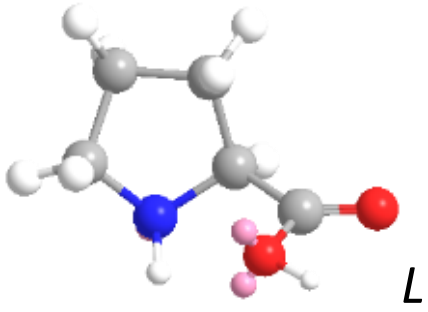 *L* | 4.20±0.01% | 3.19% |
| | 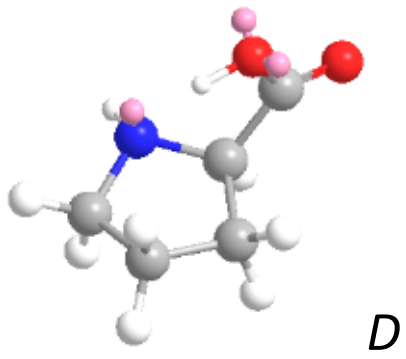 *D* | 4.07±0.02% | |
| *GO-Glycine* | 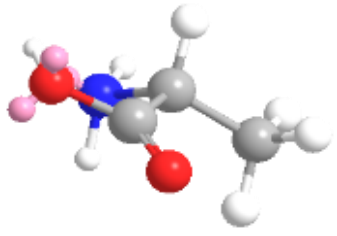 | 2.35±0.02% | - |

The content of *N* and *C* were deteremined using elemental analyzer, *N/C* is the ratio of *N* to *C* content in *GO-amino acids*, $\Delta N/C=(N/C_L-N/C_D)/(N/C_D)\times 100\%$